\documentclass[
    ,final            % use final for the camera ready runs
    ,numberedheadings % uncomment this option for numbered sections
  ]
  {aipproc}

\layoutstyle{8x11single}

\begin{document}

\title{Testing Born's Rule in Quantum Mechanics with a\\ Triple Slit Experiment}


\author{Urbasi Sinha}{
  address={Institute for Quantum Computing, University of Waterloo, 200 University Ave W,\\ Waterloo, Ontario N2L~3G1, Canada},
  email={usinha@iqc.ca}
}
\author{Christophe Couteau}{
  address={Institute for Quantum Computing, University of Waterloo, 200 University Ave W,\\ Waterloo, Ontario N2L~3G1, Canada}}
\author{Zachari Medendorp}{
  address={Institute for Quantum Computing, University of Waterloo, 200 University Ave W,\\ Waterloo, Ontario N2L~3G1, Canada}}
\author{Immo S{\"o}llner}{
  address={Institute for Quantum Computing, University of Waterloo, 200 University Ave W,\\ Waterloo, Ontario N2L~3G1, Canada},
  altaddress={Institut f\"ur Experimentalphysik, Universit\"at Innsbruck, Technikerstrasse 25, 6020 Innsbruck, Austria}}
\author{Raymond Laflamme}{
  address={Institute for Quantum Computing, University of Waterloo, 200 University Ave W,\\ Waterloo, Ontario N2L~3G1, Canada},
  altaddress={Perimeter Institute for Theoretical Physics, 31 Caroline St, Waterloo, Ontario N2L 2Y5, Canada}}
\author{Rafael Sorkin}{
  address={Department of Physics, Syracuse University, Syracuse, NY 13244-1130},
  altaddress={Perimeter Institute for Theoretical Physics, 31 Caroline St, Waterloo, Ontario N2L 2Y5, Canada}}
\author{Gregor Weihs}{
  address={Institute for Quantum Computing, University of Waterloo, 200 University Ave W,\\ Waterloo, Ontario N2L~3G1, Canada},
  altaddress={Institut f\"ur Experimentalphysik, Universit\"at Innsbruck, Technikerstrasse 25, 6020 Innsbruck, Austria},
  email={gregor.weihs@uibk.ac.at}
}

\begin{abstract}
    In Mod. Phys. Lett.~A \textbf{9,} 3119 (1994), one of us (R.D.S) investigated a formulation of quantum mechanics as a generalized measure theory. Quantum mechanics computes probabilities from the absolute squares of complex amplitudes, and the resulting interference violates the (Kolmogorov) sum rule expressing the additivity of probabilities of mutually exclusive events. However, there is a higher order sum rule that quantum mechanics does obey, involving the probabilities of three mutually exclusive possibilities. We could imagine a yet more general theory by assuming that it violates the next higher sum rule. In this paper, we report results from an ongoing experiment that sets out to test the validity of this second sum rule by measuring the interference patterns produced by three slits and all the possible combinations of those slits being open or closed. We use attenuated laser light combined with single photon counting to confirm the particle character of the measured light.
\end{abstract}

\maketitle

\section{Introduction and Motivation}
Quantum Mechanics has been one of the most successful tools in the history of Physics. It has revolutionized Modern Physics and helped explain many phenomena. However, in spite of all its successes, there are still some gaps in our understanding of the subject and there may be more to it than meets the eye. This makes it very important to have experimental verifications of all the fundamental postulates of Quantum Mechanics. In this paper, we aim to test Born's interpretation of probability \cite{Born26a}, which states that if a quantum mechanical state is specified by the wavefunction $\psi (r,t)$  \cite{Schrodinger26}, then the probability $p(\mathbf r,t)$ that a particle lies in the volume element $d^{3}r$ located at $\mathbf r$ and at time $t$, is given by:
\begin{equation}
    p(\mathbf r,t) ) = \psi^{*}(\mathbf r,t) \psi(\mathbf r,t) d^{3}r = |\psi (\mathbf r,t)|^{2} d^{3}r
\end{equation}

Although this definition of probability has been assumed to be true in describing several experimental results, no experiment has ever been performed to specifically test this definition alone. Already in his Nobel lecture in 1954, Born raised the issue of proving his postulate. Yet, 54 years have passed without there being a dedicated attempt at such a direct experimental verification, although the overwhelming majority of experiments indirectly verify the postulate when they show results that obey quantum mechanics. In this paper, we report the results of ongoing experiment that directly tests Born's rule.

\section{The 3-slit experiment}
In Ref.~\cite{Sorkin94a}, one of us (R.D.S) proposed a triple slit experiment motivated by the ``sum over histories'' approach to Quantum Mechanics.  According to this approach, Quantum theory differs from classical mechanics not so much in its kinematics, but in its dynamics, which is stochastic rather than deterministic. But if it differs from deterministic theories, it also differs from previous stochastic theories through the new phenomenon of {\it interference}. Although the quantum type of randomness is thus non-classical, the formalism closely resembles that of classical probability theory when expressed in terms of a sum over histories. Each set A of histories is associated with a non-negative real number $p_A=|A|$ called the ``quantum measure'', and this measure can in certain circumstances be interpreted as a probability (but not in all circumstances because of the failure of the classical sum rules as described below). It is this measure (or the corresponding probability) that enters the sum rules we are concerned with.Details of the quantum measure theory following a sum over histories approach can be found in \cite{Sorkin94a,Sorkin97a}.

Interference expresses a deviation from the classical additivity of the probabilities of mutually exclusive events.  This additivity can be expressed as a ``sum rule'' $I=0$ which says that the interference between arbitrary pairs of alternatives vanishes.  In fact, however, one can define a whole hierarchy of interference terms and corresponding sum-rules as given by the following equations.  They measure not only pairwise interference, but also higher types involving three or more alternatives, types which could in principle exist, but which quantum mechanics does not recognize.

\begin{equation}\label{zero}
    I_A = p_A
\end{equation}

\begin{equation}\label{one}
    I_{AB} = p_{AB} - p_A - p_B
\end{equation}

\begin{equation}\label{two}
    I_{ABC} = p_{ABC} - p_{AB} - p_{BC} - p_{CA} + p_A + p_B + p_C
\end{equation}

Equations (\ref{zero}), (\ref{one}), and (\ref{two}) refer to the zeroth, first, and second sum rule respectively. Here, $p_{ABC}$ means the probability of the disjoint union of the sets A, B, and C.  A physical system in which such probability terms appear would be a system with three classes of paths \cite{Weihs96a}, for example three slits A, B and C in an opaque aperture.  For particles incident on the slits, $p_A$ would refer to the probability of a particle being detected at a chosen detector position having traveled through slit A and $p_B$ and $p_C$ would refer to similar probabilities through slits B and C.

The zeroth sum rule needs to be violated ($ I_A \neq 0 $) for a non-trivial measure.  If the first sum rule holds, i.e. $I_{AB} = 0$, it leads to regular probability theory for example for classical stochastic processes. Violation of the first sum rule ($ I_{AB} \neq 0$) is consistent with Quantum Mechanics. A sum rule always entails that the higher ones in the hierarchy hold. However, since the first sum rule is violated in Quantum Mechanical systems, one needs to go on to check the second sum rule. In known systems, triadditivity of mutually exclusive probabilities is true i.e., the second sum rule holds, $ I_{ABC} = 0$. This follows from algebra as shown below and is based on the assumption that Born's rule holds.

\begin{eqnarray}\label{algebra}
  p_{ABC} &=& | \psi_A + \psi_B + \psi_C |^{2} \nonumber \\ &=& | \psi_A |^2 + | \psi_B |^2 + | \psi_C |^2 + \psi_{A}^{*} \psi_B +  \psi_{B}^{*} \psi_A +  \psi_{B}^{*} \psi_C +  \psi_{C}^{*} \psi_B +  \psi_{A}^{*} \psi_C +  \psi_{C}^{*} \psi_A \nonumber \\&=& p_{A} + p_{B} + p_{C} + I_{AB} + I_{BC} + I_{CA} \nonumber \\&=& p_{A} + p_{B} + p_{C} + (p_{AB} - p_A - p_B) + (p_{BC} - p_B - p_C) + (p_{CA} - p_C - p_A) \nonumber \\&=& p_{AB} + p_{BC} + p_{CA} - p_A - p_B - p_C \\\Rightarrow I_{ABC} &\equiv& p_{ABC} - p_{AB} - p_{BC} - p_{CA} + p_A + p_B + p_C =0
\end{eqnarray}

\begin{figure}[b]
  \includegraphics[width=0.5\textwidth]{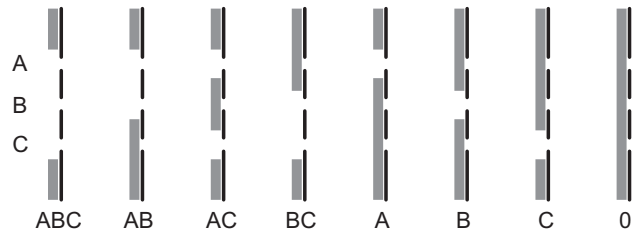}
  \caption{Pictorial representation of how the different probability terms are measured. The leftmost configuration has all slits open, whereas the rightmost has all three slits blocked. The black bars represent the slits, which are never changed or moved throughout the experiment. The thick grey bars represent the opening mask, which will is moved in order to make different combinations of openings overlap with the slits, thus switching between the different combinations of open and closed slits.}
  \label{fig:slitsandopenings}
\end{figure}

If however, there is a higher order correction to Born's rule (however small that correction might be), equation (\ref{algebra}) will lead to a violation of the second sum rule. The triple slit experiment proposes to test the second sum rule, or in more physical language, to look for a possible ``three way interference'' beyond the pairwise interference seen in quantum mechanics. For this purpose we define a quantity $\epsilon$ as
\begin{equation}\label{epsilon}
  \epsilon = p_{ABC} - p_{AB} - p_{BC} - p_{CA} + p_A + p_B + p_C -p_0.
\end{equation}
Figure~\ref{fig:slitsandopenings} shows how the various probabilities could be measured in a triple slit configuration. As opposed to the ideal formulation where empty sets have zero measure, we need to provide for a non-zero $p_0$, the background probability of detecting particles when all paths are closed. This takes care of any experimental background, such as detector dark counts. For better comparison between possible realizations of such an experiment, we further define a normalized variant of $\epsilon$ called $\rho$,
\begin{eqnarray}
    \rho &=& \frac {\epsilon}{\delta}\mbox{, where} \\
    \delta  &=&  | I_{AB} | + | I_{BC} | + | I_{CA} | \nonumber \\
    &=& | p_{AB} - p_A - p_B + p_0| + | p_{BC} - p_B - p_C + p_0 | + | p_{CA} - p_C - p_A + p_0 |.
\end{eqnarray}
Since $\delta$ is a measure of the regular interference contrast, $\rho$ can be seen as the ratio of the violation of the second sum rule versus the expected violation of the first sum rule. (If $\delta=0$ then $\epsilon=0$ trivially, and we really are not dealing with quantum behavior at all, but only classical probabilities.) In the following sections we will describe how we implemented the measurements of all the terms that compose $\rho$ and analyze our results.

\subsection{Making the slits}

\begin{figure}[b]
  \includegraphics[height=.25\textheight]{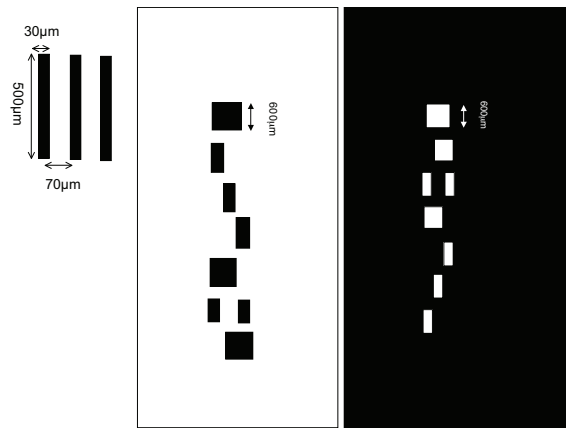}
  \caption{Different ways of measuring the eight intensities. The LHS shows a schematic of a 3 slit pattern. In the center, the first blocking scheme is demonstrated, in which the slits are blocked according to the terms being measured. The whole glass plate is thus transparent with only the blocking portions opaque. The RHS shows the second blocking scheme in which the slits are opened up as needed on a glass plate which is completely opaque except for the unblocking openings.}
  \label{blocking}
\end{figure}

Our first step in designing the experiment was to find a way to reliably block and unblock the slits, which we expected to be very close together, so that simple shutters wouldn't work. Therefore we decided to use a set of two plates, one containing the slit pattern and one containing patters to block or unblock the slits. The slits were fabricated by etching them on some material which covered a glass plate. The portion of the material which had the slits etched in would be transparent to light and the rest of the glass plate which was still covered would be opaque. However, not all materials exhibit the same degree of opacity to infra-red light and this leads to spurious transmission through portions of the glass plate which should be opaque in theory.

Various types of materials were used for etching the slits and each modification led to a decrease in spurious transmission through the glass plate. At first, a photo-emulsion plate was used which had a spurious transmission of around 5\%. This was followed by a glass plate with a chromium layer on top. This had a spurious transmission of around 3\%. The plate currently in use has an aluminium layer of 500~nm thickness on top. Aluminium is known to have a very high absorption coefficient for infrared light and this led to a spurious transmission of less than 0.1\%.

The blocking patterns were etched on a different glass plate covered with the same material as the first glass plate. Figure~\ref{blocking} shows an example of a set of blocking patterns which would give rise to the eight intensities corresponding to the probability terms related to the 3-slit open, 2-slit open and 1-slit open configurations as discussed in the previous section.

Another way of achieving the eight intensities would be to open up the right number and position of slits instead of blocking them off. This is also shown in Figure~\ref{blocking} and leads to a big change in the appearance of the second glass plate. In the first instance, when the slits were being blocked for the different cases, the rest of the glass plate was transparent and only the portions which were being used for blocking off the slits in the first glass plate were opaque to light. This led to spurious effects as a lot of light was being let through the glass plate this way, which caused background features in the diffraction patterns. However, with the second design, the whole plate was covered with the opaque material and only portions which were being used to open up slits allowed light to go through, thus leading to diminishing background effects.

\subsection{The experimental set-up}

\begin{figure}
  \includegraphics[width=.6\textwidth]{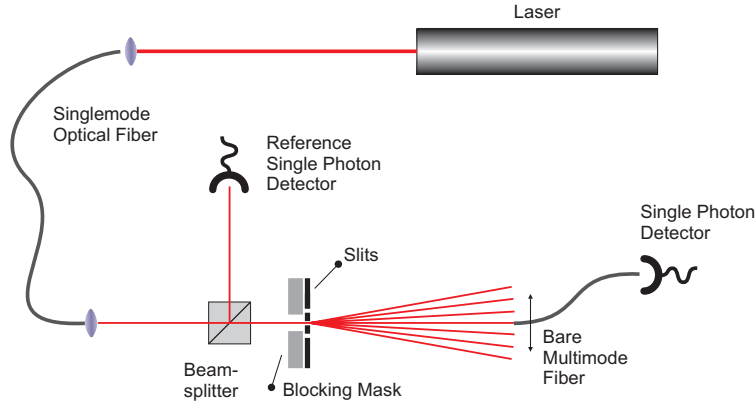}
  \caption{Schematic of experimental set-up}
  \label{slitsetup}
\end{figure}

\begin{figure}
  \includegraphics[width=.6\textwidth]{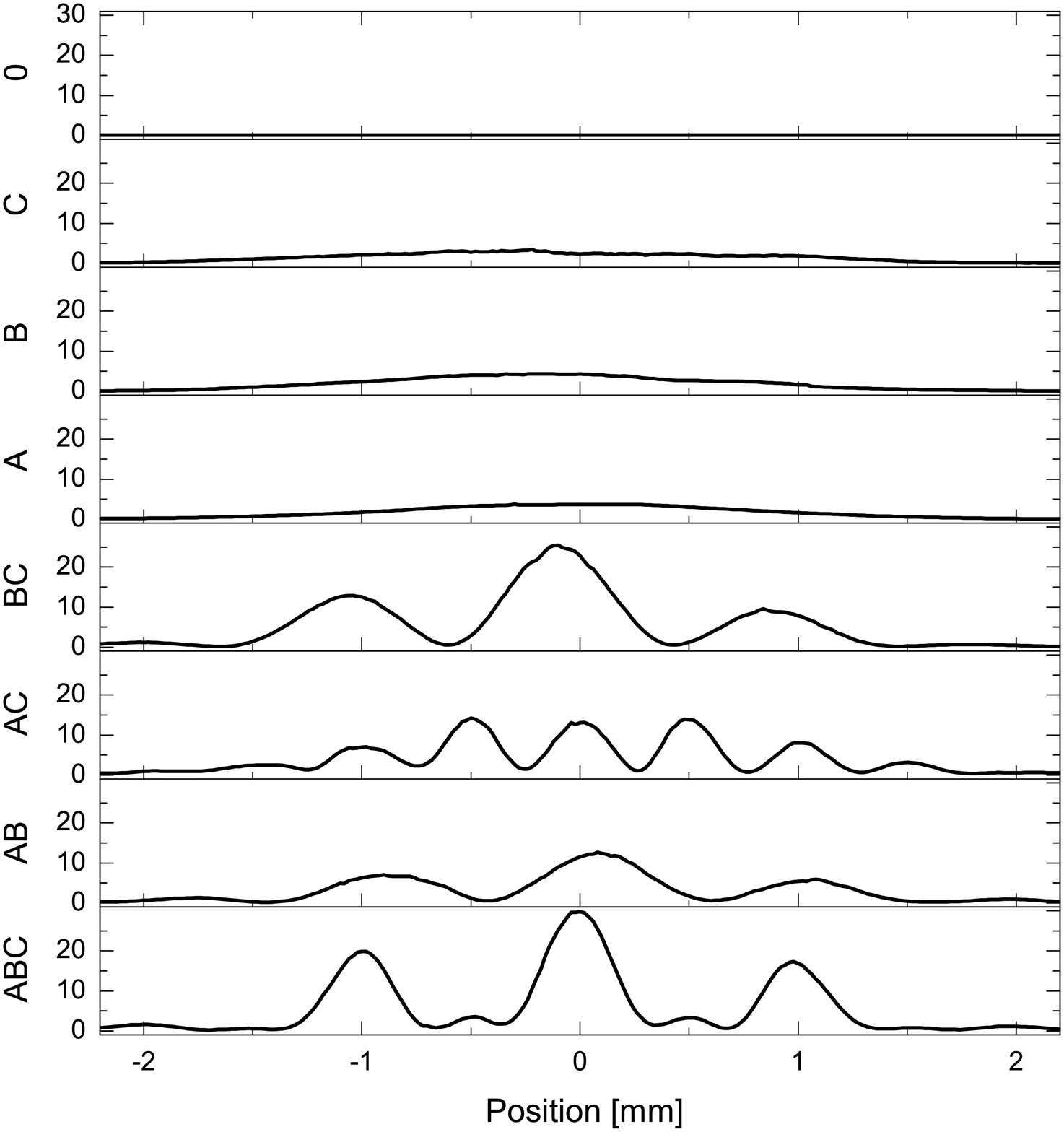}
  \caption{Diffraction patterns of the eight combinations of open and closed slits including all slits closed (``0''), measured using a He-Ne laser. The vertical axis is in units of 1000 photocounts.}
  \label{diff}
\end{figure}

Figure~\ref{slitsetup} shows a schematic of the complete experimental set-up. The He-Ne laser beam passes through an arrangement of mirrors and collimators before being incident on a 50/50 beam splitter. In the near future we will replace the laser by a heralded single photon source \cite{Bocquillon08}. The beam then splits into two, one of the beams is used as  a reference arm for measuring fluctuations in laser power whereas the other beam is incident on the glass plate, which has the slit pattern etched on it. The beam height and waist is adjusted so that it is incident on a set of three slits, the slits being centered on the beam. There is another glass plate in front which has the corresponding blocking designs on it such that one can measure the seven probabilities in equation~(\ref{two}). The slit plate remains stationary whereas the blocking plate is moved up and down in front of the slits to yield the various combinations of opened slits needed to measure the seven probabilities. As mentioned above, in our experimental set-up, we also measure an eighth probability which corresponds to all three slits being closed in order to account for dark counts and any background light. Figure~\ref{fig:slitsandopenings} shows this pictorially. There is a horizontal microscope (not shown in Figure~\ref{slitsetup}) for initial alignment between the slits and the corresponding openings. A multi-mode optical fiber is placed at a point in the diffraction pattern and connected to an avalanche photo-diode (APD) which measures the photon counts corresponding to the various probabilities. Using a single photon detector confirms the particle character of light at the detection level. The optical fiber can be moved to different positions in the diffraction pattern in order to obtain the value of $\rho$ at different positions in the pattern. Figure~\ref{diff} shows a measurement of the eight diffraction patterns corresponding to the eight configurations of open and closed slits as required by equation~(\ref{epsilon}).

\begin{figure}
  \includegraphics[width=.6\textwidth]{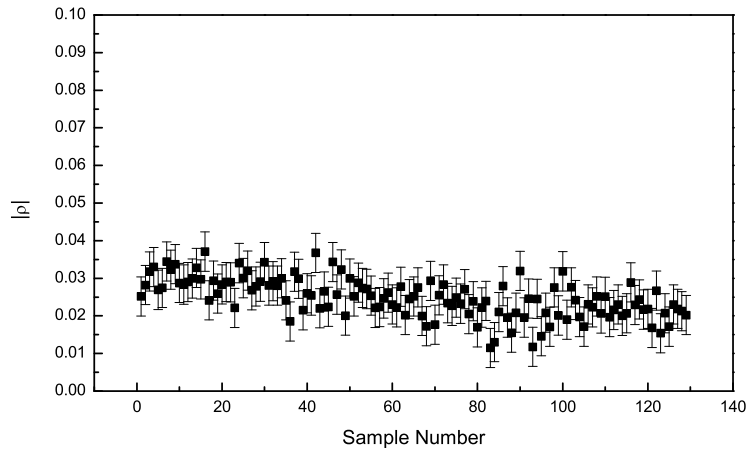}
  \caption{Overnight measurement of $\rho$. Each data point corresponds to approximately 5 min of total measurement time. A slight drift of the mean is visible. The error bars are the size of the standard deviation of the $\rho$ values.}
  \label{fig:rhodata}
\end{figure}

\subsection{Results}
In a null experiment like ours, where we try to prove the existence or absence of an effect, proper analysis of possible sources of errors is of utmost importance. It is essential to have a good estimation of both random errors in experimental quantities as well as  possible potential sources of systematic errors. For each error mechanism we calculate within the framework of some accepted theory how much of a deviation from the ideally expected value they will cause. Drifts in time or with repetition can often be corrected by better stabilization of the apparatus, but any errors that do not change in time can only be characterized by additional measurements.

We have measured $\rho$ for various detector points using a He-Ne laser. Initially, the value of $\rho$ showed strong variations with time and this was solved by having better temperature control in the lab and also by enclosing the set-up in a black box so that it is not affected by stray photons in the lab. Fig.~\ref{fig:rhodata} shows a recent overnight run which involved measuring $\rho$ around hundred times at a position near the center of the diffraction pattern. Only a slight drift in the mean can be discerned. The typical value of $\rho$ is in the range of $10^{-2} \pm 10^{-3}$. The random error is the standard error of the mean of $\rho$ and obviously it is too small to explain the deviation of the mean of $\rho$ from zero. Next we analyze some systematic errors which may affect our experiment to see if these can be big enough to explain the deviation of $\rho$ from the zero expected from Born's rule.

\section{Analysis of some possible sources of systematic errors}
By virtue of the definition of the measured quantity $\epsilon$ (or its normalized variant $\rho$) some potential sources of errors do not play a role. For example, it is unimportant whether the three slits in the aperture have the same size, shape, open transmission, or  closed light leakage. However, in the current set-up we are measuring the eight different combinations of open and closed slits using a blocking mechanism that does not block individual slits but by changing a global unblocking mask. Also, the measurements of the different combinations occur sequentially, which makes the experiment prone to the effects of fluctuations and drifts. In the following we will analyze the effects of three systematic error mechanisms, power drifts or uneven mask transmission, spurious mask transmission combined with misalignment, and detector nonlinearities.

The power of a light source is never perfectly stable and the fact that we measure the eight individual combinations at different times leads to a difference in the total energy received by a certain aperture combination over the time interval it is being measured for. Since in practice we don't know how the power will change, and because we may choose a random order of our measurements we can effectively convert this systematic drift into a random error. Conversely, if in the experiment we found that the power was indeed drifting slowly in one direction, then randomization of the measurement sequence would mitigate a non-zero mean.

\begin{figure}
  \includegraphics[width=0.8\textwidth]{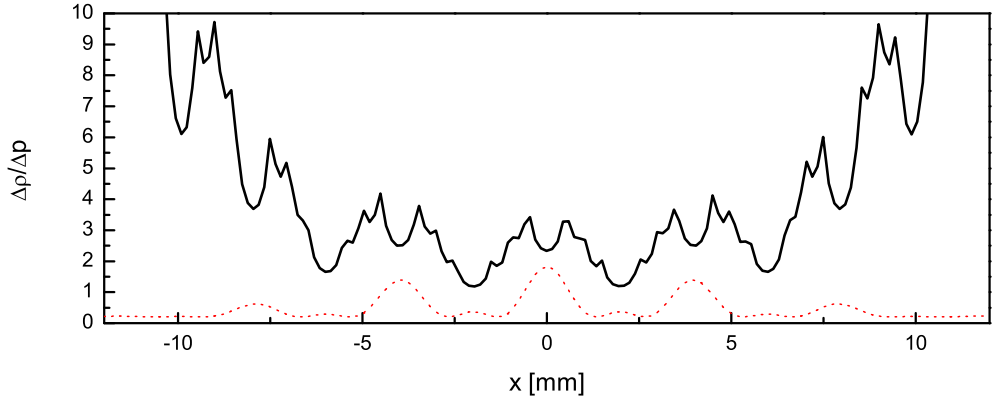}
  \caption{Fluctuation $\Delta\rho$ of $\rho$ caused by fluctuating source power $\Delta p$ (solid line). The horizontal axis is the spatial coordinate in the far field of the three slits. The dotted line shows a scaled three-slit diffraction pattern as a position reference.}
  \label{fig:powererror}
\end{figure}

Let us therefore assume a stationary mean power $P$ and a constant level of fluctuations $\Delta P$ around that power for an averaging time that is equal to the time we take to measure one of the eight combinations. Let the relative fluctuation $\Delta p = \Delta P/P$. Using Gaussian error propagation, the fluctuation $\Delta\rho$ of $\rho$, whose quantum theoretical mean is zero, is then given by
\begin{eqnarray}
  (\Delta\rho)^2  &=& \frac{1}{\delta^2} \left[
    P_{ABC}^2 +
    (1+s_{BC}\rho) P_{BC}^2 +
    (1+s_{AC}\rho) P_{AC}^2 +
    (1+s_{AB}\rho) P_{AB}^2 + \right. \\ \nonumber && \;\;\;\;\;\;\;
    (1+(s_{BC}+s_{AC})\rho) P_{C}^2 +
    (1+(s_{BC}+s_{AB})\rho) P_{B}^2 +
    (1+(s_{AC}+s_{AB})\rho) P_{A}^2 + \\ \nonumber && \;\;\;\;\;\;\;
    \left.(1+(s_{BC}+s_{AC}+s_{AB})\rho) P_{0}^2
  \right] (\Delta p)^2, \label{eq:powererror}
\end{eqnarray}
where the quantities $s$ are the signs of the binary interference terms that appear in $\delta$, e.g. $s_{AB}=\mathrm{sign}(I(A,B))$. Fig.~\ref{fig:powererror} shows a plot of $\Delta \rho/\Delta p$ as a function of the position in the diffraction pattern. The curve has divergences wherever $\delta$ has a zero. These are the only points that have to be avoided. Otherwise, the relative power stability of the source translates with factors close to unity into the relative error of $\rho$.

Obviously, Eq.~\ref{eq:powererror} is also exactly the formula for the propagation of independent random errors of any origin in the measurements, if they are all of the same relative magnitude. However, if we use a photon counting technique, then the random error of each measurement follows from the Poissonian distribution of the photocounts. In this case, the (relative) random error of $P_x$ is proportional to $1/\sqrt{P_x}$, where $x$ is any of the eight combinations. As a consequence the random error of $\rho$ will be proportional to the same expression with all the $P_x^2$ replaced by $P_x$.

While it appears that drifting or fluctuating power can be mitigated, a worse problem is that in our realization of the unblocking of slits every pattern could potentially have slightly different transmission. Possible reasons for this are dirt, or incomplete etching of the metal layer, or inhomogeneities in the glass substrate or the antireflection coating layers. In order to avoid any of these detrimental possibilities the next implementation of the slits will be air slits in a steel membrane.

\begin{figure}
  \includegraphics[width=0.8\textwidth]{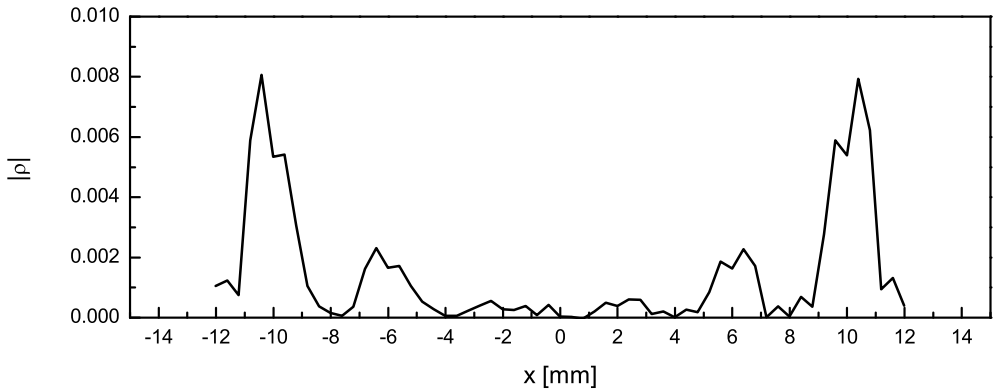}
  \caption{Value of $\rho$ in the diffraction pattern of three slits for the following set of parameters: $30\;\mu\mathrm m$ slit size, $100\;\mu\mathrm m$ slit separation, 800~nm wavelength, $100\;\mu\mathrm m$ opening size, 5\% unwanted mask transmission, and a set of displacements of the blocking mask uniformly chosen at random from the interval [0,$10\;\mu\mathrm m$]}
  \label{fig:blockingerror}
\end{figure}

As a second source of systematic errors we have identified the unwanted transmission of supposedly opaque parts of the slit and blocking mask. This by itself would not cause a non-zero $\rho$, but combined with small errors in the alignment of the blocking mask it will yield aperture transmission functions that are not simply always the same open and closed slit, but in every one of the eight combinations we have a particular aperture transmission function. If the slits were openings in a perfectly opaque mask there would be no effect, since they are not being moved between the measurements of different combinations. In practice, we found that all our earlier masks had a few percent of unwanted transmission as opposed to the current one which has an unwanted transmission smaller than 0.1\%. Fig.~\ref{fig:blockingerror} shows the results of a simulation assuming the parameters of the current mask,  which seems to be good enough to avoid this kind of systematic error at the cur  rent level of precision.

\begin{figure}
  \includegraphics[width=0.9\textwidth]{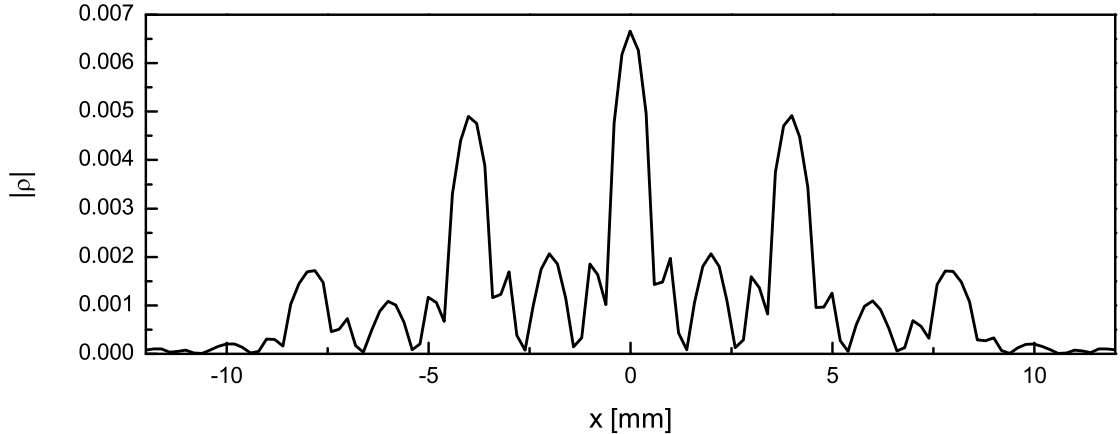}
  \caption{Value of $\rho$ in the diffraction pattern of three slits for a 0.5\% nonlinear detector, where the ratio between the maximum detected power and the minimum detector power is 100.}
  \label{fig:nlerror}
\end{figure}

Finally, there is a source of systematic error, which is intrinsically linked to the actual objective of this measurement. We set out to check the validity of Born's rule, that probabilities are given by absolute squares of amplitudes. Yet, any real detector will have some nonlinearity. In a counting measurement the effect of dead-time will limit the linearity severely, even at relatively low average count rates. A typical specification for an optical power meter is 0.5\% nonlinearity within a given measurement range. The measurement of all eight combinations involves a large dynamic range. From the background intensity to the maximum with all three slits open, this could be as much as six orders of magnitude. Fig.~\ref{fig:nlerror} shows that 1\% nonlinearity translates into a non-zero value of $\rho$ of up to 0.007. For the measurements shown above the mean count rate was about 80,000 counts per second. Given a specified dead time of our detector of 50~ns, we expect the deviation from linearity to be about 0.4\% and the resulting apparent value of $\rho \approx 0.003$.

All of these systematics are potential contributors to a non-zero mean $\rho$. From the above calculations and our efforts to stabilize the incident power and improvements in the mask properties, we conclude that while detector nonlinearities may have contributed something, the main source of systematic error must be the inhomogeneities in the unblocking mask. Hopefully air slits will bring a significant improvement.

\section{Discussion, conclusion and future work}
In this experiment, we have attempted to test Born's rule for probabilities. This is a null experiment but due to experimental inaccuracies, we have measured a value of $\rho$ which is about $10^{-2} \pm 10^{-3}$. We have analyzed some major sources of systematic errors that could affect our experiments and we will try to reduce their influence in future implementations. Further, we plan replace the random laser source by a heralded single photon source \cite{Bocquillon08}. This will ensure the particle nature of light both during emission and detection and give us the advantage that we can count the exact number of particles entering the experiment. At this point we don't know of any other experiment that has tried to test Born's rule using three-path interference, therefore we cannot judge how well we are doing. However, our collaborators \cite{Cory08a} are undertaking an interferometric experiment using neutrons, which will perform the test in a completely different   system. These two approaches are complementary and help us in our quest to estimate the extent of the validity of the Born interpretation of the wavefunction.

\section{Acknowledgements}
Research at IQC and Perimeter Institute was funded in part by the Government of Canada through NSERC and by the Province of Ontario through MRI. Research at IQC was also funded in part by CIFAR. This research was partly supported by NSF grant PHY-0404646. U.S. thanks Aninda Sinha for useful discussions.

\bibliographystyle{aipproc}   % if natbib is available
\bibliography{confproc}

\end{document}